\documentclass[aps,jcp,preprint,superscriptaddress]{revtex4}
\usepackage[utf8]{inputenc}

\usepackage{amsmath, amssymb}
\usepackage{tabularx,booktabs,array,dcolumn}
\usepackage{setspace}
\usepackage{graphicx,psfrag,subfigure}
\usepackage{color}

\usepackage[all]{xy}

\usepackage{enumitem}
\usepackage{mdframed}
\usepackage{xcolor}
\usepackage{rotating}

\newcommand{\mx}[1]{\boldsymbol{#1}}
\newcommand{\bos}[1]{\boldsymbol{#1}}
\newcommand{\mr}[1]{\mathrm{#1}}

\newcommand{\pd}[2]{\frac{\partial #1}{\partial #2}}
\newcommand{\cm}{cm$^{-1}$}

\def\Eh{$E_\text{h}$}

\def\tr{^{\text{T}}}

\def\mel{m_\mr{e}}

\def\tlg{\tilde g}

\def\ra0{\mx{a}}

\def\som{Supplementary Material}

\def\htwo{\scalebox{0.5}{2}}

\def\el{\text{el}}
\def\htwo{\hat{H}^\text{(2)}}
\def\ttwo{\hat{T}^\text{(2)}}
\def\hbo{\hat{H}^\text{(0)}}

\def\mmx{\mathcal{A}}

\def\tlg{\tilde{g}}

\def\hel{\hat{H}_\text{el}}

\def\naux{N_\text{aux}}

\def\massnuc{m_{\text{n}}}

\def\tmassvib{\tilde{m}_{\text{vib}}}
\def\tmassrot{\tilde{m}_{\text{rot}}}

\def\XdSup{X\ {^2\Sigma}_{\text{u}}^+}

\def\paperone{Paper~I}

\begin{document}

\title{%
Non-adiabatic mass-correction functions
and rovibrational states of $^4$He$_2^+$ ($X\ ^2\Sigma_\text{u}^+$)
}

\author{Edit M\'atyus}
\email{matyus@ chem.elte.hu}
\affiliation{Institute of Chemistry, E\"otv\"os Lor\'and University, P\'azm\'any P\'eter s\'et\'any 1/A, Budapest, H-1117, Hungary}
\date{\today}

\begin{abstract}
\noindent %
The mass-correction functions in the second-order non-adiabatic Hamiltonian
are computed for the $^4$He$^+_2$ molecular ion using 
the variational method, floating explicitly correlated
Gaussian functions, and a general coordinate-transformation 
formalism.
When non-adiabatic rovibrational energy levels are computed using these 
(coordinate-dependent) mass-correction functions and a highly accurate
potential energy and diagonal Born--Oppenheimer correction curve, 
significantly improved theoretical results are 
obtained for the nine rotational and two rovibrational 
intervals known from high-resolution spectroscopy experiments.
\end{abstract}

\maketitle

\clearpage
\section{Introduction}
\noindent %
Precision spectroscopy measurements challenge theoretical and computational methodologies
to go much beyond the current state of the art and to study various possible subtle effects, 
which have been overlooked or considered to be negligible in the past. 
The search for ``small effects'' is usually triggered by a disagreement between
the most accurate experimental and the best possible theoretical results, when this disagreement
is larger than the error bars on the experimental and computed datasets. 
A famous example for a decades-long, hand-in-hand race of theory and experiment is
the dissociation energy of molecular hydrogen \cite{UbMe09,SpJuUbMe11}, which started at the advent of 
quantum mechanics and has reached by now a level of sophistication when 
experts propose to use the hydrogen molecule's spectral lines to test fundamental physical
constants and search for new physical theories \cite{BeHoAgDeScMe18,AlDrSaUbEi18}, 
thereby cultivating a renaissance for molecular physics.

The present work is motivated by the sizeable disagreement 
between the recently measured rotational intervals of $^4$He$_2^+$ \cite{SeJaMe16}
and adiabatic rotational-vibrational computations using a
highly accurate potential energy curve, 
including the diagonal Born--Oppenheimer corrections (DBOC),
obtained with variationally optimized
floating explicitly correlated Gaussian (fECG) functions in Ref.~\cite{TuPaAd12}.
The deviation of the experimental and computational results 
increases with the rotational quantum number, 
$N^+$, the energy of which is measured from the $N^+=1$
rotational state (in the electronic and vibrational ground state). 
The deviation becomes as large as $0.069$~\cm\ (2.1~GHz) for 
the $\tilde\nu(N^+=19)-\tilde\nu(N^+=1)$ interval,
which is almost twice as large as the estimated relativistic and radiative corrections.
Semeria, Jansen, and Merkt proposed \cite{SeJaMe16} that 
the neglect of the coordinate-dependence of the effective masses,
which would account for non-adiabatic effects, 
could be responsible for the ``missing'' part of the deviation 
between experiment and theory. 

The second-order non-adiabatic Hamiltonian for a single isolated electronic state,
which includes a kinetic-energy-correction term that can be written as 
a correction to the mass tensor, has
been discovered and re-discovered in various contexts in the 
past \cite{FiKi64,HeAs66,BuMo77,BuMo80,BuMcMo77,WeLi93,AFHaUm94,BrReRu96,HeOg98,Og98book,Sch01H2p,Sch01H2O,SaJeOg05,BaSaOdOg05,PaKo08,PaKo09,HoSzFrRePeTy11,PaKo12,PaSpTe07,prx17}---for a hopefully nearly complete
summary see the Introduction of Ref.~\cite{Ma18nonad} (henceforth \paperone). 
These derivations and discoveries have
remained somewhat isolated yet, and the rigorous computation of the mass-correction tensor is 
rarely carried out, whereas the DBOC, which appears at the same order in the Hamiltonian,
is almost routinely computed for a variety of poly-electronic and 
poly-atomic molecules \cite{HaYaSc86,HaLe96,IoAmHa96,VaSh03}. 
In the present work, we use the general curvilinear formalism
and our in-house developed variational fECG code \cite{Ma18nonad} to obtain the rigorous mass-correction
functions for the $^4$He$_2^+$ molecular ion in its $X\ ^2\Sigma_\text{u}^+$ ground electronic state,
and compute its non-adiabatic rovibrational bound and long-lived resonance states.

\clearpage
\section{Theoretical and computational details \label{ch:theo}}
\noindent %
In \paperone\ we have re-written the second-order non-adiabatic Hamiltonian, $\htwo$, of Ref.~\cite{PaSpTe07}, 
using a general curvilinear coordinate formalism and considered the time-independent Schrödinger equation:
\begin{align}
  \left(%
    -\frac{1}{2}\tlg^{-1/2}\partial_\mu\tlg^{1/2}\tilde{G}^{\mu\nu}\partial_\nu 
    + U + V
  \right)
    \phi = E \phi
  \label{eq:rovibsolve}
\end{align}
where $U$ is the diagonal Born--Oppenheimer correction to the BO potential energy, $V$, 
and the effective $\tilde{G}^{\mu\nu}$ tensor 
was written as:
\begin{align}
  \tilde{G}^{\mu\nu}
  &=
  \frac{1}{{\massnuc}}
  g^{\mu\alpha}
  (%
    \mx{S}
    \mx{I}_6
    \mx{S}\tr
  )_{\alpha\beta}  
  g^{\beta\nu}
  -
  \frac{1}{{\massnuc}^2}
  g^{\mu\alpha}
  (%
    \mx{S}
    \bar{\mmx}
    \mx{S}\tr
  )_{\alpha\beta}  
  g^{\beta\nu}
  \nonumber \\
  &=
  \left[%
    \frac{1}{{\massnuc}}
    \delta^\mu\,_\beta
    -
    \frac{1}{{\massnuc}^2}  
    \mmx^\mu\,_\beta\ 
  \right]
  g^{\beta\nu} .
  \label{eq:shc}
\end{align}
In Eq.~(\ref{eq:shc}), 
$g^{\alpha\beta}$ is the $(\alpha,\beta)$th element of the 
contravariant metric tensor of the curvilinear coordinates and we
defined the transformation matrix
\begin{align}
  \mx{S}
  =
  \mx{J}\tr\mx{O}.
  \label{eq:trfomx}
\end{align}
In Eq.~(\ref{eq:trfomx}) $\mx{J}$ is the Jacobian matrix for the curvilinear coordinates
and $\mx{O}$ is a rotation matrix which connects the laboratory frame and the frame
of the atomic nuclei used to compute the elements of the $\bar{\mmx}$ mass-correction tensor 
from electronic-structure theory. For a body-fixed frame selected for the atomic nuclei 
($R_{ia},\ i=1,\ldots,N$ and $a=x(1),y(2),z(3)$)
$\bar{\mmx}$ is computed as \footnote{We use the expanded-, $(ia)$, and 
condensed-index, ${I=3(i-1)+a}$,
labelling introduced in \paperone\ in an interchangeable manner.}
\begin{align}
  \bar{\mmx}_{ia,jb}(\mx{R})
  &=
  2\langle%
    \partial_{R_{ia}}\psi|
    (\hat{H}_\el-E)^{-1}(1-\hat{P})
    \partial_{R_{jb}}\psi
  \rangle_\el 
  \nonumber \\
  &\approx
  2\sum_{n=1}^{\naux}
  \sum_{m=1}^{\naux}
  \langle %
    \partial_{R_{ia}}\psi| f_n 
  \rangle_\el
  \left(\bos{F}^{-1}\right)_{nm}
  \langle %
    f_m |\partial_{R_{jb}}\psi
  \rangle_\el 
  \label{eq:Amxauxbas}
\end{align}
with 
\begin{align}
  (\bos{F})_{nm}(E)
  =
  \langle 
    f_n |
    (\hel - E) (1-\hat{P})
    | f_m
  \rangle_\el,
  \label{eq:Fmx}
\end{align}
and we have introduced the $\lbrace f_n,n=1,\ldots\naux\rbrace$ auxiliary basis set 
to compute the resolvent. $1-\hat{P}$ is a projector to the space orthogonal to the electronic
state $\psi$, which is an ``isolated'' eigenstate of the $\hel$ electronic Hamiltonian, 
for further details see \paperone.

Derivatives of the electronic wave functions are computed by finite differences with a step
size of $\Delta=10^{-4}$~bohr along each Cartesian directions and using the rescaling idea
for the fECG centers 
upon a small change of the nuclear positions \cite{CeKu97,Ma18nonad}.
Instead of directly computing $\mx{F}^{-1}$, we solve the system of linear equations 
for $\mx{x}_{m,j_b}$
\begin{align}
  [\mx{F}(E+\text{i}\varepsilon)\ \mx{x}_{j_b}]_m
  = 
  \langle %
    f_m |\partial_{R_{jb}}\psi
  \rangle_\el ,
  \label{eq:Fxb}
\end{align}
where instead of using $E$, we have $E+\text{i}\varepsilon$ which allows us to 
avoid the implicit or explicit inversion of a near singular matrix ($E$ is a tight variational
upper bound to an eigenvalue of the electronic Hamiltonian, $\hat{H}_\text{el}$).
For the present applications, $\text{i}\varepsilon=\text{i}10^{-3}-\text{i}10^{-5}$~E$_\text{h}$
ensured numerically stable results (mass-correction values) over the $\rho\in[1,10]$~bohr
internuclear distance interval to at least 4 significant digits 
(for $\rho>10$~bohr we have approximated the mass-correction functions with their asymptotic value, 
see Fig.~\ref{fig:masscorrhe2px}).

For homonuclear diatomic molecules, it is natural to replace
the six Cartesian coordinates with the three polar coordinates,
$(\rho,\theta,\phi)$ and the three Cartesian coordinates of the nuclear center of mass (NCM). 
This transformation and the particular expressions for the effective 
$\tilde{G}^{\mu\nu}$ tensor have been worked out in detail in \paperone, 
so we only repeat the final expression (``.'' means ``0'' in the matrix):
\begin{align}
  \tilde{G}^{\mu\nu}
  &=
  {\tiny
  \left(%
    \begin{array}{@{}c@{}c@{}c | c@{}c@{}c @{}}
      \frac{2}{\massnuc}\left[1-\frac{\mmx^\rho\,_\rho}{\massnuc}\right] &  . &  . & . & . & . \\
       . & \frac{2}{\massnuc}\left[1-\frac{\mmx^\theta\,_\theta}{\massnuc}\right] \frac{1}{\rho^2} &  . & . & . & . \\
       . & . & \frac{2}{\massnuc}\left[1-\frac{\mmx^\phi\,_\phi}{{\massnuc}}\right]\frac{1}{\rho^2\sin^2\theta} & . & . & . \\
       \cline{1-6}
       . &  . &  . & \quad\quad\quad\quad  & \quad\quad  &   \\
       . &  . &  . & \multicolumn{3}{c}{\frac{1}{2\massnuc}\left[1-\frac{\mmx^\text{NCM}\ _\text{NCM}}{{\massnuc}}\right]\mx{I}_3} \\
       . &  . &  . & \quad\quad\quad\quad  & \quad\quad &  \\       
     \end{array}
 \right)   
 }
 \label{eq:diatGeff}
\end{align}
It is important to re-iterate that not only in the  BO, $\hbo$, but also in $\htwo$,
the translation of the nuclear center of mass (always) exactly separates
from the translationally invariant (TI), \emph{i.e.,} rotational-vibrational, 
part of the problem (in $\tilde{G}^{\mu\nu}$ the TI-NCM block is rigorously zero) \cite{prx17},
hence we consider only the rovibrational part by subtracting the translational kinetic energy of
the center of mass (which also gains a correction term in $\ttwo$).
Furthermore, 
the kinetic energy operator of diatomic molecules ($\hat{T}^{(0)}$ or $\hat{T}^{(2)}$)
does not contain any mixed derivatives between the radial and angular coordinates
(note that the TI block in Eq.~(\ref{eq:diatGeff}) is diagonal).
Hence, the angular part of the second-order rovibrational Hamiltonian 
can be integrated with the $Y_{JM}(\theta,\phi)$ 
spherical harmonic functions (similarly to the standard solution of diatomics with $\hat{T}_{\text{rv}}^{(0)}$), 
and we are left with the numerical solution of the radial equation:
\begin{align}
  &\left(%
    -\frac{1}{2}\frac{1}{\rho^2}
    \pd{}{\rho}
    \rho^2 
    \frac{2}{\massnuc}
    \left[%
      1-\frac{\mmx^\rho\,_\rho}{\massnuc}
    \right] 
    \pd{}{\rho}
  \right.
  \nonumber \\
  &+\left. 
    \frac{J(J+1)}{\rho^2}
    \frac{1}{\massnuc}
    \left[%
      1 - \frac{\mmx^\Omega\,_\Omega}{\massnuc}
    \right]
    + U(\rho) + V(\rho)
  \right)
  \varphi_J(\rho) 
  =E_J \varphi_J(\rho) 
  \label{eq:diatrad1}
\end{align}
where $\varphi_J$ is normalized using the volume element $\rho^2\text{d}\rho$.
Instead of solving Eq.~(\ref{eq:diatrad1}), 
we proceed similarly to Pachucki and Komasa \cite{PaKo09} and use the operator identity
\begin{align}
 \frac{1}{\rho^2}\pd{}{\rho} \rho^2 X(\rho) \pd{}{\rho} 
 =
 \frac{1}{\rho}\pd{}{\rho} X(\rho) \pd{}{\rho} \rho 
 - \frac{1}{\rho} \pd{X(\rho)}{\rho} 
\end{align}
to obtain
\begin{align}
  &\left(%
    -\pd{}{\rho} 
    \frac{1}{\massnuc}
    \left[%
      1-\frac{\mmx^\rho\,_\rho}{\massnuc}
    \right] 
    \pd{}{\rho}
  \right.
  \nonumber \\
  &\left. 
   -\frac{1}{\rho}\frac{1}{\massnuc^2} \pd{\mmx^{\rho}\,_\rho}{\rho}
   +\frac{J(J+1)}{\rho^2}
    \frac{1}{\massnuc}
    \left[%
      1 - \frac{\mmx^\Omega\,_\Omega}{\massnuc}
    \right]
    +
    U(\rho) + V(\rho)
  \right)
  \chi_J(\rho) 
  =E_J \chi_J(\rho),  
  \label{eq:diatrad2}
\end{align}
where $\chi_J(\rho)=\rho \varphi_J(\rho)$ and $\chi_J$ is normalized with the volume element $\text{d}\rho$.
To obtain rovibrational energies (including non-adiabatic corrections) and wave functions,
we solve Eq.~(\ref{eq:diatrad2}) using the discrete variable representation (DVR) and 
associated Laguerre polynomials, $L_n^{(\alpha)}$ with $\alpha=2$ for the
radial (vibrational) degree of freedom.
The DVR points are scaled to an $[R_\text{min},R_\text{max}]$ interval. 
The $n$ number of DVR points and functions, as well as $R_\text{min}$ and $R_\text{max}$ are 
determined as convergence parameters, and their typical value is around $n=300-1000$, 
$R_\text{min}=1$~bohr and $R_\text{max}=30-100$~bohr.
Finally, we mention that 
the term including the $\partial\mathcal{A}^{\rho}\,_\rho/\partial\rho$ derivative
has an almost negligible effect on the computed rotational-vibrational
states ($<0.001$~\cm).

\subsection{Mass-correction curves and non-adiabatic rovibrational energies\label{ch:he2pnum}}
For the rovibrational computations, we used the highly accurate PES and DBOC curves
computed by Tung, Pavanello, and Adamowicz \cite{TuPaAd12}. 
The mass-correction functions have been evaluated in the present work using a modified
version of our in-house developed pre-Born--Oppenheimer (preBO) code 
\cite{MaHuMuRe11a,MaHuMuRe11b,MaRe12,Ma13,Ma18},
see also related developments in Refs.~\cite{SiMaRe13,SiMaRe14,SiMaRe15,MuMaRe18}.
The particular computational methodology for the mass-correction tensor is explained in Paper~I, 
and the numerical details for the present system are described in the following paragraphs.

\begin{figure}[h!]
  \includegraphics[scale=1.03]{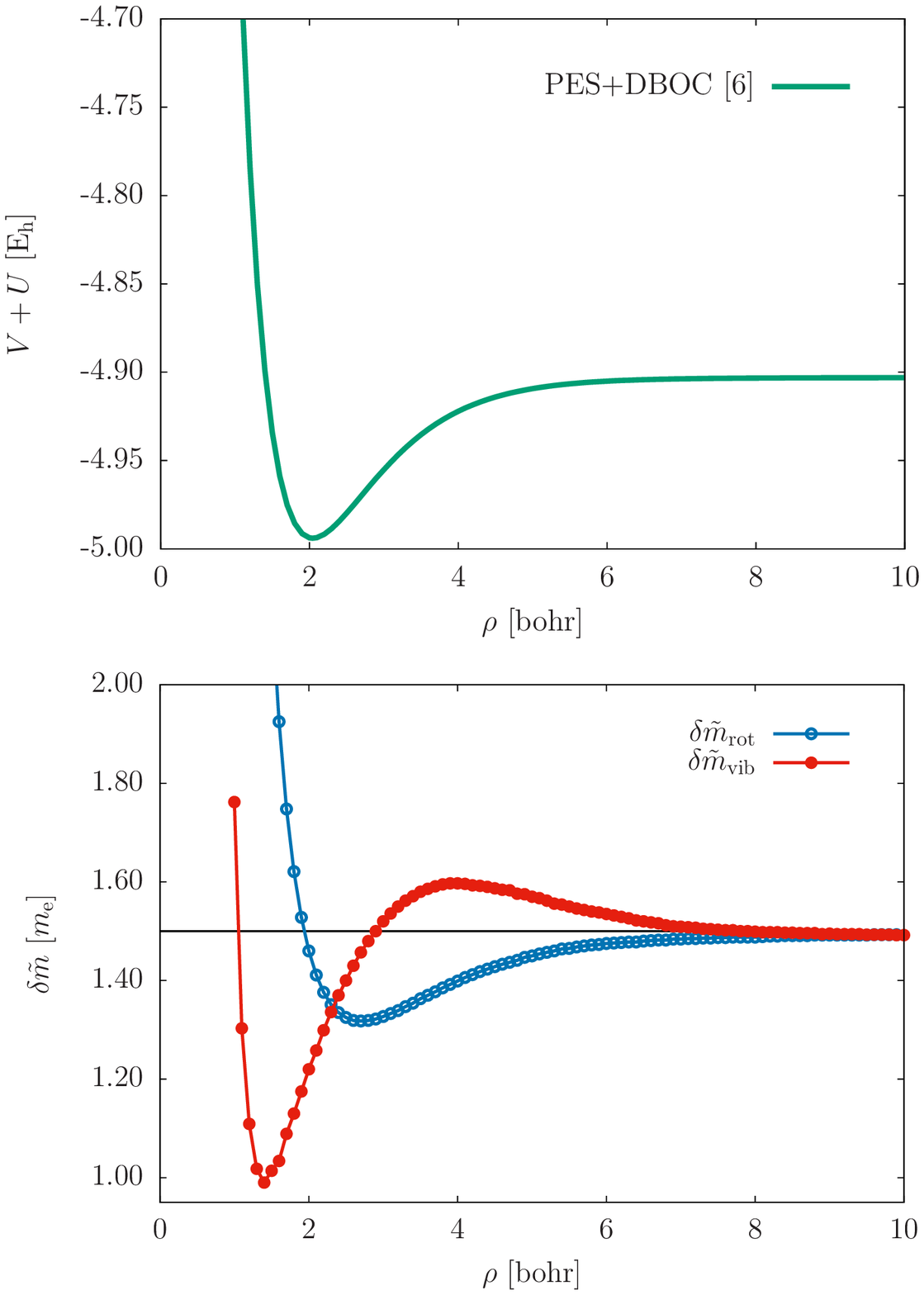}\\
  \caption{%
    $^4$He$_2^+$ molecular ion in its
    $\XdSup$ (ground) electronic state:
    rotational, 
    $\delta\tilde{m}_\text{rot}=\mmx^\Omega\,_\Omega$ 
    and
    vibrational, 
    $\delta\tilde{m}_\text{vib}=\mmx^\rho\,_\rho$    
    mass-correction functions
    computed in the present work.
    The thin line takes the constant value of $3/2=1.5\ m_\text{e}$, which corresponds to an 
    equal share of the three electrons between
    the two atomic nuclei. The adiabatic PES curve (upper plot) 
    is reproduced using the points
    of Ref.~\cite{TuPaAd12} to show the value and change of the mass-correction 
    functions with respect to the PES valley.
    \label{fig:masscorrhe2px}
    }
\end{figure}

We optimized the  floating ECG basis function parameters as well as 
the single parameter in the spin function of three electrons coupled to 
a doublet state \cite{SuVaBook98} by minimizing the electronic energy at 
the internuclear distance $\rho=2$~bohr. 
The electronic energy was converged within $10\ \mu$\Eh\ ($\sim 2.2$~\cm) in comparison 
with earlier work \cite{CeRy95,TuPaAd12}.
Starting out from the optimized electronic basis set corresponding to $\rho=2$~bohr,
we have increased (and decreased) the internuclear distance step by step with
$\Delta \rho=0.1$~bohr to cover the $\rho\in[1,10]$~bohr interval. 
At each step, we have rescaled the basis set origins \cite{CeKu97} 
and performed an entire refinement cycle of the non-linear parameters.
This rescaled and refined basis set was used for making the next step along
the series of nuclear configurations (the refinement was necessary in order to 
maintain the accuracy of the results after making several nuclear displacements).
The resulting basis sets provided us with an electronic energy accurate within ca. 10--15~$\mu$\Eh\ 
over the $\rho\in[1,10]$~bohr interval.
We have carried out a variety of convergence tests for the mass correction functions
with respect to 
(a) the convergence of the electronic state wave function, $\psi$;
and 
(b) the size of the auxiliary basis set used to 
express the resolvent, Eqs.~(\ref{eq:Amxauxbas})--(\ref{eq:Fmx}).
A quick assessment of the auxiliary basis set was possible based on the observation 
that the mass correction terms were inaccurate if the insertion
of the truncated resolution of identity 
in the derivative overlaps gave inaccurate results (see Sec.~IVB of \paperone).
The auxiliary basis set of increasing size 
was compiled from the basis sets optimized for the electronic
wave functions
at neighboring nuclear configurations: $\rho$, $\rho\pm0.1$~bohr,
$\rho\pm0.2$~bohr, $\rho\pm0.3$~bohr. 
The computed mass-correction functions for the radial and angular degrees
of freedom, $\delta\tilde{m}_\text{vib}=\mmx^\rho\,_\rho$ and 
$\delta\tilde{m}_\text{rot}=\mmx^\Omega\,_\Omega$, respectively, 
are visualized in Figure~\ref{fig:masscorrhe2px} (the numerical values are deposited in the \som). 

The first nine rotational intervals measured from 
the $N^+=1$ rotational state (in the electronic and vibrational ground state) of $^4$He$_2^+$ are given
in Table~\ref{tab:rotcompHe2p} (in what follows, we adopt the notation of Ref.~\cite{SeJaMe16}
and use $N^+$ for the rotational angular 
momentum quantum number of He$_2^+$). For a direct comparison with the experimentally observed 
rotational intervals, we added to the non-adiabatic transition energies 
the estimated relativistic and radiative
corrections \cite{TuPaAd12}. According to the explanation in Ref.~\cite{SeJaMe16}
these corrections were calculated by scaling the relativistic and radiative corrections of the
rotational states of H$_2$~\cite{PiJe09} by a factor of 3.9.

\begin{figure}[h]
\includegraphics[scale=1.]{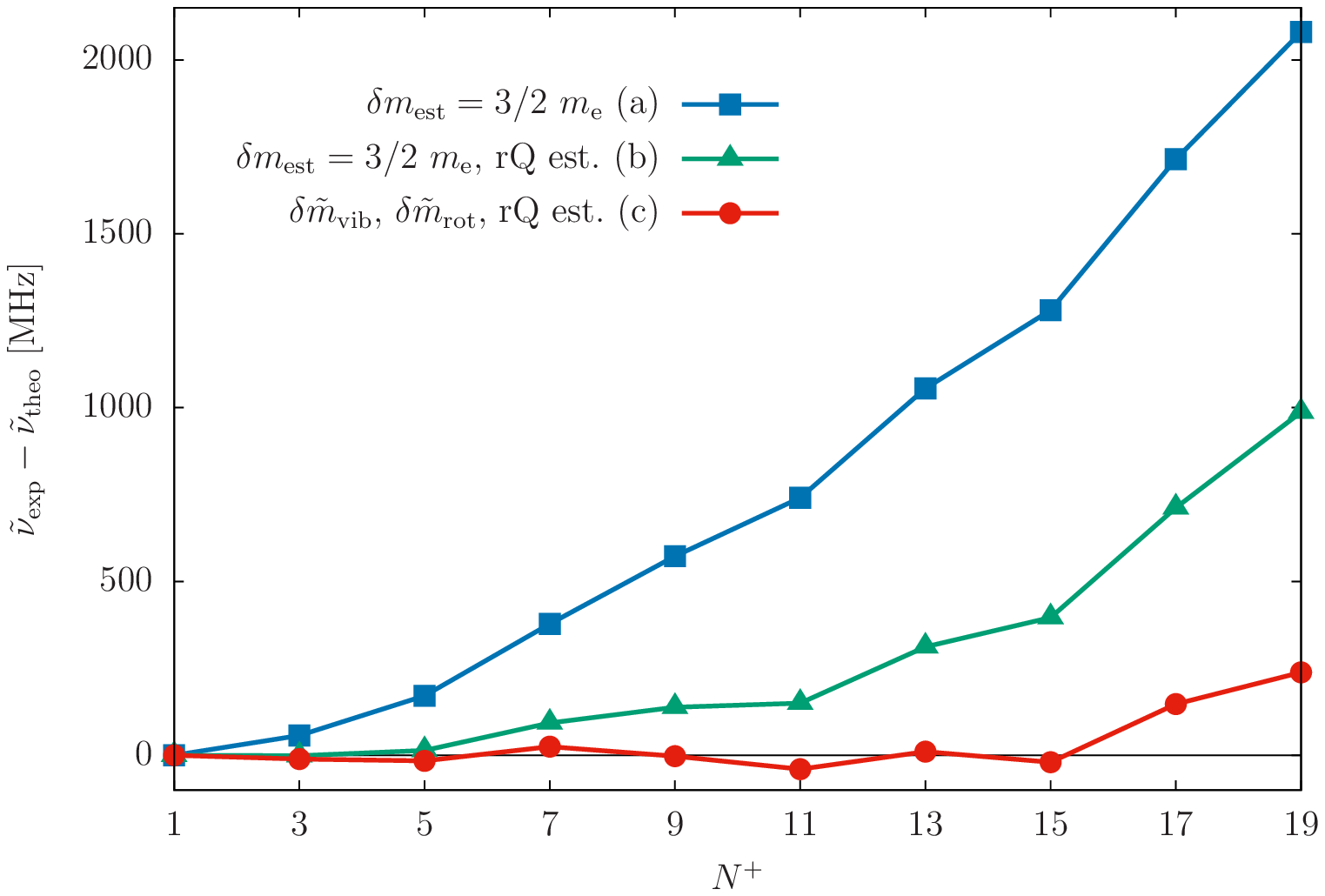}
\caption{%
  Deviation of the experimentally measured \cite{SeJaMe16} 
  and computed rotational transitions of the
  $^4$He$_2^+$ molecular ion in its ground electronic and vibrational 
  state ($X\ ^2\Sigma_{\text{u}}^+$, $v=0$).
  The rotational intervals were computed using 
  the PES and DBOC curves of Ref.~\cite{TuPaAd12}
  and 
  (a) the $\delta m_\text{est} =3/2$ constant-mass correction model also used in Ref.~\cite{TuPaAd12} (blue squares);
  (b) the same as (a) but including the relativistic and radiative estimates \cite{TuPaAd12}, see text and Ref.~\cite{SeJaMe16} for details (green triangles); and
  (c) using the rigorous mass-correction functions computed in the present work and including the relativistic
  and radiative estimates (red circles).
  \label{fig:He2pdevexp}
}
\end{figure}

The deviation of the computed rotational intervals
from experiment is visualized in Figure~\ref{fig:He2pdevexp}. 
In particular for high rotational angular
momentum values the explicit inclusion of the rotational and vibrational mass curves
significantly improves the agreement with the experimental results 
in comparison with the na\"ive choice of an effective 
mass including the 3/2 mass of an electron (an equal share of the mass of the three electrons
between the two atomic nuclei).

In order to attach error bars to our computational results, 
we can think about the following possible sources of errors and inaccuracies.
By repeating the evaluation of the mass-correction functions using an increasing number of
electronic and auxiliary basis functions, we estimate that the 
$\mmx^\rho\,_\rho$ and $\mmx^\Omega\,_\Omega$ values are converged to within about 0.1--0.5~\% 
(see also Paper~I for the numerical and convergence properties of the mass-correction functions of
H$_2^+$). 
A 1~\% change of the rotational mass correction function ($1\pm0.01$~$\mmx^{\Omega}\,_\Omega$) 
results in a $\pm 0.005$~\cm\ change in the $\Delta\tilde\nu(N^+=19)-\tilde\nu(N^+=1)$ interval
(for lower $N^+$ upper states the change is smaller).
So, the uncertainty of the non-adiabatic rotational intervals with respect to 
the enlargement of the electronic and auxiliary basis sets is probably smaller than this value.
The pure rotational intervals measured in Ref.~\cite{SeJaMe16} 
are not sensitive to inaccuracies in the vibrational mass correction function 
(a $\pm 50$~\% change of $\mmx^{\rho}\,_\rho$ results in a $\mp 0.001$~\cm\ change in
$\Delta\tilde\nu=\tilde\nu(N^+=19)-\tilde\nu(N^+=1)$). 

It is much more difficult for us to make an assessment about the error in the estimated 
relativistic and radiative corrections, and firm error bounds could be made by
explicitly computing these corrections for the $^4$He$_2^+$ molecular ion.

To finish with the discussion of the numerical results, we comment on the two earlier 
$^4$He$_2^+$ measurements \cite{MaAsNoLoPeBu76,CaPyKn95}.
Ref.~\cite{MaAsNoLoPeBu76} reported predissociative features of $^4$He$_2^+$. 
In the hope of being able to identify (and assign) the reported peaks
we computed ``all'' bound and long-lived resonance states of $^4$He$_2^+$
using the rigorous mass-correction curves computed in this work and the highly accurate PES and DBOC
curves of Ref.~\cite{TuPaAd12}.
Due to the low resolution and relatively large uncertainty
of the experimental energy positions of \cite{MaAsNoLoPeBu76}, 
we refrain here from giving any suggestions for the assignment,
and just deposit the computed non-adiabatic rovibrational energies 
(and non-adiabatic corrections) in the \som.

Using the high-resolution microwave electronic transitions reported in Ref.~\cite{CaPyKn95},
two rovibrational energy intervals of the ground-electronic state can be deduced.
These intervals were computed in Ref.~\cite{TuPaAd12} with 
the constant $\delta m=3/2\ \mel$ mass correction model,
and we have also computed them with the coordinate-dependent mass functions  (see Table~\ref{tab:He2provib}).
The $(v,N^+)=(23,3)-(23,1)$ interval was obtained in excellent agreement 
with the experiments already with the constant mass model~\cite{TuPaAd12}
and a similarly excellent agreement is observed using the rigorous mass-correction functions.
For the $(v,N^+)=(23,3)-(22,5)$ interval the authors of Ref.~\cite{TuPaAd12}
observed a larger, 0.012~\cm, deviation from experiment. 
When using the rigorous mass functions, 
this deviation is reduced to $-0.003$~\cm, which is
probably in the order of magnitude of the relativistic and radiative corrections, 
which (as explained earlier) should be addressed in future work.

\clearpage

\begin{table}[h]
\caption{%
  Comparison of the experimentally measured and computed rotational intervals, 
  in \cm, of the
  $^4$He$_2^+$ ion in its electronic and vibrational ground state ($\XdSup$, $v=0$).
  In all computations we used the highly accurate PES and DBOC curves 
  computed by Tung \emph{et al.} \cite{TuPaAd12}. 
  The nuclear mass for the $^4$He nucleus was $\massnuc=7\,294.299\,536\,3\ m_\text{e}$
  (same as in Ref.~\cite{TuPaAd12}) and $1~\mathrm{E}_\mathrm{h}=219\,474.6313\,702$~\cm\ \cite{codata06}.
  \label{tab:rotcompHe2p}
}
\begin{tabular}{@{}c@{\ \ \ }c@{\ \ \ } r@{\ }l c r@{\ }l c r@{}r@{}}
\cline{1-10}\\[-0.4cm]
\cline{1-10}\\[-0.4cm]
$N^+$	&	
$\tilde{\nu}_\text{exp}$ $^\text{a}$ & 
$\tilde{\nu}_\text{na-est}$ &
($\delta\tilde{\nu}_\text{na-est}$) $^\text{b}$ &&
$\tilde{\nu}_\text{na}$ &
($\delta\tilde{\nu}_\text{na}$) $^\text{c}$ &&
$\tilde{\nu}_\text{na-rQest}$ &
($\delta\tilde{\nu}_\text{na,rQ-est}$) $^\text{d}$ \\
\cline{3-4}
\cline{6-7} 
\cline{9-10}\\[-0.4cm]
&  &
\multicolumn{2}{c}{$[\delta m_\text{est}=3/2m_\text{e}]$} &&
\multicolumn{2}{c}{$[\delta\tmassrot,\delta\tmassvib]$} &&
\multicolumn{2}{c}{$[\delta\tmassrot,\delta\tmassvib,\text{rQ est.}]$}\\
\cline{1-10}\\[-0.4cm]
1	&	0		&	0	 &		&&	0	&		&&	0	&		\\
3	&	70.9379		&	70.936	 &	(0.002)	&&	70.936	&	(0.002)	&&	70.938	&	(0.000)	\\
5	&	198.3647	&	198.359	 &	(0.006)	&&	198.360	&	(0.005)	&&	198.365	&	($-0.001$)	\\
7	&	381.8346	&	381.822	 &	(0.013)	&&	381.824	&	(0.010)	&&	381.834	&	(0.001)	\\
9	&	620.7021	&	620.683	 &	(0.019)	&&	620.688	&	(0.014)	&&	620.702	&	(0.000)	\\
11	&	914.1367	&	914.112	 &	(0.025)	&&	914.118	&	(0.018)	&&	914.138	&	($-0.001$)	\\
13	&	1261.1242	&	1261.089 &	(0.035)	&&	1261.099 &	(0.025)	&&	1261.124 &	(0.000)	\\
15	&	1660.4627	&	1660.420 &	(0.043)	&&	1660.434 &	(0.029)	&&	1660.463 &	($-0.001$)	\\
17	&	2110.7932	&	2110.736 &	(0.057)	&&	2110.755 &	(0.038)	&&	2110.788 &	(0.005)	\\
19	&	2610.5744	&	2610.505 &	(0.069)	&&	2610.530 &	(0.044)	&&	2610.566 &	(0.008)	\\
\cline{1-10}\\[-0.4cm]
\cline{1-10}
\end{tabular}
\begin{flushleft}
  $^\text{a}$: Experimental results taken from Table I of Ref.~\cite{SeJaMe16}. \\
  $^\text{b}$: 
    Rotational intervals computed 
    with constant effective nuclear masses corresponding to the 
    qualitative reasoning that the electrons follow the nuclei: 
    $m_\text{est}=\massnuc+\delta m_\text{est}$ (both for rotations and vibrations). 
    These effective masses were used also in Ref.~\cite{TuPaAd12} to compute 
    rotation-vibration energy levels. \\
  $^\text{c}$: 
     Rotational intervals computed with
     the rigorous mass-correction functions computed 
     in the present work (see Figure~\ref{fig:masscorrhe2px}). \\
  $^\text{d}$: Rotational transitions obtained in $^\text{c}$ 
    corrected with the estimated relativistic and radiative effects for each rotational state
    according to Ref.~\cite{TuPaAd12} (see also text and Ref.~\cite{SeJaMe16} for details). \\
\end{flushleft}
\end{table}

\begin{table}[h]
\caption{%
  Comparison of rovibrational transition energies of $^4$He$_2^+$
  derived from experiment \cite{CaPyKn95} and computed 
  with the adiabatic potential energy curve of Ref.~\cite{TuPaAd12}
  and constant effective masses ($\tilde{\nu}_\text{na-est}$) 
  as well as with the rigorous mass-correction curves ($\tilde{\nu}_\text{na}$)
  computed in this work.
  \label{tab:He2provib}
}
  \begin{tabular}{@{}c@{\ \ }c@{\ \ }c@{\ \ \ }c c@{\ }l@{\ \ \ }c@{\ }l@{}}
    \cline{1-8}\\[-0.4cm]
    \cline{1-8}\\[-0.4cm]
    $(v',{N^+}')$ & 
    $(v'',{N^+}'')$ & 
    $\tilde{\nu}_\text{exp}$$^\text{a}$ &
    \multicolumn{2}{c}{$\tilde{\nu}_\text{na-est}$ 
    ($\delta\tilde{\nu}_\text{na-est}$) $^\text{b}$} &&
    \multicolumn{2}{c}{$\tilde{\nu}_\text{na}$ 
    ($\delta\tilde{\nu}_\text{na}$) $^\text{c}$} \\
    \cline{4-5}
    \cline{7-8}
    \\[-0.4cm]
    & & & 
    \multicolumn{2}{c}{$[\delta m_\text{est}=3/2m_\text{e}]$} &&
    \multicolumn{2}{c}{$[\delta\tmassrot,\delta\tmassvib]$} \\
    \cline{1-8}
    \\[-0.4cm]
    (23,3) & 
    (22,5) &    
    5.248 &
    5.260 &
    $(0.012)$ &&
    5.250 &
    \multicolumn{1}{r}{$(-0.002)$} \\
    (23,3) & 
    (23,1) &
    2.001 &
    2.002 &
    $(0.001)$ &&
    1.999 &
    \multicolumn{1}{r}{$(0.002)$} \\    
    \cline{1-8}\\[-0.4cm]
    \cline{1-8}\\[-0.4cm]
  \end{tabular}
\begin{flushleft}
  $^\text{a}$: Experimental results of Ref.~\cite{CaPyKn95}. \\
  $^\text{b}$: See footnote (b) to Table~\ref{tab:rotcompHe2p}. \\
  $^\text{c}$: See footnote (c) to Table~\ref{tab:rotcompHe2p}. \\
\end{flushleft}
\end{table}

\clearpage
\section{Summary and conclusions}
\noindent%
We have computed the rigorous rotational and vibrational mass-correction functions
in the second-order non-adiabatic Hamiltonian \cite{PaSpTe07}
for the $^4$He$_2^+$ molecular ion in its $X\ ^2\Sigma_\text{u}^+$
(ground) electronic state. The computations have been carried out using
the variational method and floating explicitly correlated Gaussian functions
and a general curvilinear formalism of the second-order kinetic-energy operator including the
mass-correction tensor developed in Ref~\cite{Ma18nonad}. 
Solution of the rovibrational time-independent Schrödinger equation
including the computed, rigorous, non-adiabatic mass-correction functions results in a promising improvement 
in comparison with experimental results. If
estimates for the relativistic and radiative corrections \cite{TuPaAd12} 
are also included, deviation of the rotational intervals, $\tilde\nu(N^+)-\tilde\nu(1)$ 
(with $N^+=3,5,$\ldots,19) is reduced to less than 100 MHz for up to $N^+=15$, which increases to 240~MHz 
for $N^+=19$.
In order to be able to attach rigorous error bars (and a possible improvement
for $N^+\geq 17)$ to the theoretical results, it will be necessary
to explicitly compute the relativistic and radiative corrections for $^4$He$_2^+$.
At the present stage, we may conclude that the highly accurate PES and DBOC points
of Ref.~\cite{TuPaAd12} and the series of high-precision rotational intervals measured 
in \cite{SeJaMe16} made it possible to study and identify subtle non-adiabatic effects
in one of the simplest poly-electronic molecule (molecular ion).

\vspace{1cm}
\noindent\textbf{Note added at the revision stage} \\
After submission of this work, 
we became aware of new measurements which determine 
rotational intervals for the first vibrational excitation ($v=1$)
and the vibrational fundamental frequency of $^4$He$_2^+$ \cite{JaSeMe18}.
The increasing experimental dataset will allow a more detailed study of the fine
interplay of small effects earlier neglected in the theoretical description of this molecular ion.

\vspace{1cm}
\noindent\textbf{Supplementary Material}\\ 
The Supplementary Material contains 
1) non-adiabatic rovibrational energies obtained with the rigorous mass-correction functions;
2) deviation of the rovibrational energies computed with the non-adiabatic mass-correction functions 
and constant masses (nuclear plus 3/2 electron mass);
3) non-adiabatic mass correction values to the rotational and vibrational degrees of freedom.

\vspace{1cm}
\noindent\textbf{Acknowledgment}\\ 
Financial support of the Swiss National Science Foundation through
a PROMYS Grant (no. IZ11Z0\_166525) is gratefully acknowledged. 
The author is thankful to Frédéric Merkt for discussions 
about the experimental observations and relativistic and radiative effects 
of the He$_2^+$ molecular ion.
We also thank Ludwik Adamowicz and co-workers for computing and sharing their highly accurate
potential energy and DBOC points through Ref.~\cite{TuPaAd12}, which made it
possible to study fine non-adiabatic phenomena.

\clearpage

\end{document}